\documentstyle{article}

\font\tenrm=cmr10
\font\tenit=cmti10
\font\elevenbf=cmbx10 scaled\magstep 1
\font\elevenrm=cmr10 scaled\magstep 1
\font\elevenit=cmti10 scaled\magstep 1

\font\ninerm=cmr9

\renewenvironment{thebibliography}[1]
 { \tenrm
 \baselineskip=10pt
   \begin{list}{\arabic{enumi}.}
    {\usecounter{enumi} \setlength{\parsep}{0pt}
     \setlength{\itemsep}{3pt} \settowidth{\labelwidth}{#1.}
     \sloppy
    }}{\end{list}}

\begin{document}
\begin{center}{\elevenbf
BARYONIC SYSTEMS WITH CHARM AND BOTTOM IN THE BOUND STATE 
SOLITON MODEL
\footnote{\ninerm\baselineskip=10pt
The work supported in part by the Russian Fund for Fundamental Research, grant
95-02-03868a and presented at the Workshop on Science at Japan Hadron
Facility (JHF98), KEK, Tsukuba, March 4-7, 1998}
\\}
\vglue 0.5cm
{\tenrm V.B.Kopeliovich \\}
{\tenit Institute for Nuclear Research of the Russian Academy of
Sciences,\\ 60th October Anniversary Prospect 7A, Moscow 117312 \\}
\vglue 0.6cm
\end{center}
{\rightskip=3pc
 \leftskip=3pc
 \tenrm\baselineskip=11pt
 \noindent
The binding energies of baryonic systems with baryon number $B=2, 3$ 
and $4$ possessing heavy flavor, charm or bottom, are estimated within 
the rigid 
oscillator version of the bound state approach to chiral soliton models. Two
tendencies are noted: the binding energy increases with increasing mass
of the flavor and with increasing $B$. Therefore,
the charmed or bottomed baryonic systems have more chances to be bound
than strange baryonic systems discussed previously. The flavor symmetry 
breaking in decay constants $F$ is considered which is especially important
for baryonic systems with bottom quantum number.
 \vglue 0.2cm}
 \vglue 0.1cm
\baselineskip=13pt
\elevenrm
\section{Introduction}
Many efforts have been done lately to investigate the properties of 
baryonic systems ($BS$) with nonzero strangeness, first of all the 
possibility of the existence of states stable relative to strong decays.

Recently some of the predictions of theory began to find experimental 
confirmation. The near-threshold enhancement in $\Lambda\Lambda$ system
observed in $\cite{1}$ can be interpreted as a component of $27$-plet
 obtained from the 
bound $SU(2)$ torus-like configuration with $B=2$ by means of collective
coordinates method described in \cite{2,3}.
Similar enhancement in $\Lambda N$ system has been observed many years ago
in the kaon production reaction on nucleons \cite{4} and confirmed also in
$\Lambda p$ scattering \cite{5}. It can belong to $27$-plet or to antidecuplet 
of dibaryons. The singlet $NN$ scattering state with isospin $T=1$
belongs to the $27$-plet (for review of theoretical predictions in $B=2$ 
sector see, e.g. \cite{6}). Analogous results are obtained in more 
conventional potential approach as well.

The question if the $BS$ with flavor different from $u$ and $d$ can exist, 
is more general, of course. Charm, bottom or top quantum numbers are
also of interest. Their consideration can be performed in the framework of
chiral soliton models, in particular, the bound state approach to heavy
flavors proposed in \cite{7} and developed in \cite{8}-\cite{10}.
Although charmed and bottomed $BS$ have less chances to play
some important role in astrophysics than the strange ones (it is not 
excluded, however!) their studies can be very useful for understanding
of the peculiarities of nuclear matter fragments with unusual properties.
 It might be similar to heavy quarkonia which studies were very important for 
 development and checking of $QCD$ itself.

Here the baryonic systems with heavy flavors are considered within the
rigid oscillator version of the bound state approach to strange baryons
proposed by Kaplan and Klebanov \cite{9} and used later in \cite{10}.
This model has definite advantages before collective coordinates 
quantization method when heavy flavors are included into consideration,
first of all, because of its simplicity. However, some apparent drawbacks
are present also.\\
\section{Rigid oscillator model}
 The ansatz for the chiral fields used in \cite{9,10} is:
$$ U(r,t) = R(t) U_0(r) R^{\dagger}(t), \qquad
 R(t) = A(t) S(t), \eqno (1) $$
where $U_0$ is $SU(2)$ soliton embedded into $SU(3)$ in usual way (into
left upper corner), $A(t) \in SU(2) $ describes $SU(2)$ rotations, 
$S(t) \in SU(3)$ describes 
rotations in the "charm" or "bottom" direction.
For definiteness we shall consider the extension of the $(u,d)$ $SU(2)$
Skyrme model in charmed direction, when $D$ is the field of $D$-mesons. 
But it is clear that quite similar the extension can be made in bottom 
and top direction. 
$$ S(t) = exp (i {\elevenit D} (t)),  \qquad
 {\elevenit D} (t) = \sum_{a=4,...7} D_a(t) \lambda_a, \eqno (2) $$
$\lambda_a$ are Gell-Mann matrices of $(u,d,c)$ or $(u,d,b)$
$SU(3)$ groups. The $(u,d,b)$ $SU(3)$ subgroup is quite analogous to
the $(u,d,s)$ one, for $(u,d,c)$ subgroup simple redefiniton of hypercharge
should be made.
 $D_4=(D^0+\bar{D}^0)/\sqrt{2}$, $D_5=i(D^0-\bar{D}^0)/\sqrt{2}$, etc.

After some calculation the well known Lagrangian  of the Skyrme model 
in the lowest order in field $D$ takes the form \cite{9,10}:
$$ L=-M_{cl,B}+4\Theta_{F,B} \dot{D}^{\dagger}\dot{D}-\Gamma_B(m_D^2-m_{\pi}^2)
D^{\dagger}D +i{N_cB \over 2}(D^{\dagger}\dot{D}-\dot{D}^{\dagger}D). \eqno(3)$$
Here $D$ is a doublet formed by $D^0$ and $D^-$ mesons,
and we maintained our former notation for the moment of inertia for the
rotation into "strange", "charm" or "bottom" direction $\Theta_c=
\Theta_b=\Theta_s=\Theta_{F}$.
This moment of inertia has simple analytical form for arbitrary starting
$SU(2)$ skyrmion, regardless its symmetry properties:
$$\Theta_{F,B} = {1 \over 8} \int (1-c_f) \bigl[F_{\pi}^2+{1 \over e^2}
\bigl( (\vec{d}f)^2+s_f^2(\vec{d}\alpha)^2+s_f^2s_{\alpha}^2(\vec{d}\beta)^2
\bigr) \bigr] d^3r, \eqno (4a) $$
$F_\pi$ and $e$ are the parameters of the model.
The general parametrization of the $SU(2)$ skyrmions has been used here,
$U = c_f+s_f \vec{\tau}\vec{n}$ with $n_z=c_{\alpha}$, $n_x=s_{\alpha}
c_{\beta}$, $n_y=s_{\alpha}s_{\beta}$, $s_f=sinf$, $c_f=cosf$, etc. 
For the axially symmetrical ansatz
$\beta=n\phi$, $\phi$ is the azimuthal angle, and $\Theta_{F,B}$ takes the form
drawn in \cite{11}:
$$\Theta_{F,B}= {\pi \over 4} \int (1-c_f) \bigl[ F_{\pi}^2+{1 \over e^2}
\bigl((f,f)+s_f^2(\alpha,\alpha) + {n^2 \over r^2}s_f^2s_{\alpha}^2\bigr)
\bigr] rdrdz=\frac{\Gamma_B}{4}+\Theta_{F,B}^{Sk}, \eqno (4b) $$
$(f,f)=(\partial f/\partial r)^2+(\partial f/\partial z)^2$, $r$ and $z$
being cylindrical coordinates. The quantity $\Gamma_B$ defines the contribution
of the mass term in the Lagrangian:
$$ \Gamma_B = {F^2_\pi \over 2} \int (1-c_f) d^3r. \eqno (5) $$
Numerical values of $\Theta_{F,B}$, $\Gamma_B$ and some other quantities 
are shown in Table 1 below. 

The term in $(3)$ proportional to $N_cB$ appears from the Wess-Zumino-Witten 
term in the action and is responsible, within this approach, for the splitting
between excitation energies of charm and anticharm (flavor and antiflavor
in general case) \cite{8}-\cite{10}. $N_c$ is the number of colors in 
the underlying
$QCD$, in all other cases here the index $c$ means the charm quantum number.
$B$ is the baryon number of the configuration which can be written in terms
of the functions $f, \alpha$ and $\beta$ as
$$ B = -{1 \over 2\pi^2} \int s_f^2 s_{\alpha}(\vec{\partial} f 
\vec{\partial} \alpha \vec{\partial} \beta ) d^3r, \eqno (6) $$
In other words, it is the Wronskian of the system described by 3 profiles, 
$f$, $\alpha$ and $\beta $ \cite{2}. For the axially symmetrical configuration
possessing also symmetry $z \rightarrow -z$, $B=n (f(0)-f(\infty))/\pi=n$
for configurations of lowest energy.

As it was noted in \cite{10} deviations of the field $D$ from the vacuum 
decrease with increasing mass $m_D$, as well as with increasing number of 
colors $N_c$, and the method works at any $m_D$ - for bottom and even top
quantum number also.
The zero modes quantum corrections due to rotation with the matrix $A(t)$
have the order of magnitude $N_c^{-1}$ and are not crucial but also important 
(see also section 4).\\
\section{Flavor excitation frequences}
After the canonical quantization procedure the Hamiltonian of the 
system takes the form:
$$H_B=M_{cl,B} + {1 \over 4\Theta_{F,B}} \Pi^{\dagger}\Pi + \bigl(\Gamma_B 
\bar{m}^2_D+\frac{N_c^2B^2}{16\Theta_{F,B}} \bigr) D^{\dagger}D -i 
{N_cB \over 8\Theta_{F,B}}
(D^{\dagger} \Pi- \Pi^{\dagger} D). \eqno (7) $$
$\bar{m}_D^2 = m_D^2-m_\pi^2$
The momentum $\Pi$ is canonically conjugate to variable $D$.
Eq. $(7)$ describes the oscillator-type motion of the field $D$ 
in the background formed 
by the $(u,d)$ $SU(2)$ soliton. After the diagonalization which can be done
explicitely according to \cite{9,10} the Hamiltonian can be written as
$$H_B= M_{cl,B} + \omega_{F,B} a^{\dagger} a + \bar{\omega}_{F,B} b^{\dagger} b
 + O(1/N_c) \eqno (8) $$
with $a^\dagger$, $b^\dagger$ being the operators of creation of charm and
anticharm (bottom and antibottom) quantum number, $\omega_{F,B}$ and 
$\bar{\omega}_{F,B}$ being the 
frequences of heavy flavor (antiflavor) excitation. $D$ and $\Pi$ are connected
with $a$ and $b$ in the following way \cite{9,10}:
$$ D^i= \frac{1}{\sqrt{N_cB\mu_{F,B}}}(a^i+b^{\dagger i}), \qquad
\Pi^i = \frac{\sqrt{N_cB\mu_{F,B}}}{2i}(a^i - b^{\dagger i}) \eqno (9) $$
with
$$ \mu_{F,B} =( 1 + 16 \bar{m}_D^2  \Gamma_B \Theta_{F,B}/ (N_cB)^2 )^{1/2}. $$
For the lowest states the values of $D$ are small,
$$ D \sim \bigl[4\Gamma_B\Theta_{F,B}\bar{m}_D^2 + N_c^2B^2/4 \bigr]^{-1/4}. $$
The flavor (antiflavor) excitation frequences $\omega$ and $\bar{\omega}$ are:
$$ \omega_{F,B} = \frac{N_cB}{8\Theta_{F,B}} ( \mu_{F,B} -1 ), \qquad
 \bar{\omega}_{F,B} = \frac{N_cB}{8\Theta_{F,B}} ( \mu_{F,B} +1 ) \eqno (10) $$
It should be noted that the difference 
$\bar{\omega}_{F,B}-\omega_{F,B} = N_cB/(4\Theta_{F,B})$ coincides in the 
leading order in $N_c$ with that obtained in the collective coordinates 
approach \cite{12,13}. Indeed, in the collective coordinates approach the
zero-modes energy of the soliton rotated in the $SU(3)$ configuration space
and depending on the "flavor" inertia $\Theta_{F,B}$ can be written
as:
$$ E_{rot}(\Theta_{F,B})={1 \over 4\Theta_{F,B}}\bigl[N_cB +n_{q\bar{q}}
 \bigl( N_cB +2n_{q\bar{q}} +2 -2T_r\bigr) \bigr] \eqno (11) $$
where $n_{q\bar{q}}$ is the number of additional quark-antiquark pairs 
present in the
quantized state, $N_cB+3n_{q\bar{q}}=p+2q$, $p,q$ are the numbers of indices 
in the spinor describing the $SU(3)$ irrep, $T_r=(p+n_{q\bar{q}})/2$ is the so 
called right
isospin characterizing irrep (see \cite{13} where the $B=1, n_{q\bar{q}}=0$ 
case was considered, and \cite{12} where $(11)$ was obtained for $N_c=3$).
The term proportional to $n_{q\bar{q}}N_cB$ in $(11)$ coincides with the 
difference of $\bar{\omega}_{F,B}-\omega_{F,B}$ in $(10)$. 

For the difference of the frequences of excitation in cases of $B \geq 2$ 
and $B=1$ systems we obtain:
$$\Delta \omega \simeq {\bar{m}_F \over 2} \biggl[\biggl
(\frac{\Gamma_1}{\Theta_{F,1}} \biggr)^{1/2}-
\biggl(\frac{\Gamma_B}{\Theta_{F,B}}\biggr)^{1/2}\biggr]  \eqno(12) $$
It is proportional to the heavy quark mass $m_F$
and is positive if $\Gamma_{1}/\Theta_{F,1} \geq \Gamma_B/\Theta_{F,B}$.
For $B=2,3$ it is really so. 
The characteristics of $SU(2)$ toroidal solitons with baryon numbers 
$B=2,3$ and $4$ have been calculated previously \cite{14}.
For $B=2$ they coincide with good accuracy with those given later in \cite{10}.
For greater baryon numbers some configurations of lower energy have been found
\cite{15,16}, but necessary quantities like $\Theta_{F,B}$ and $\Gamma_B$ are 
absent, still.
 
As a result, the binding 
energy of heavy flavored dibaryons, tribaryons, etc. increases in comparison
with strange flavor case, as it can be seen from 
the results of numerical estimates shown in Table 1.\\
\section{$1/N_c$ zero modes corrections and binding energies estimates}
 The $\sim 1/N_c$ zero modes quantum correction to the energies of
$BS$ can be estimated according to the expression \cite{9,10}:
$$\Delta E_{1/N_c} = {1 \over 2\Theta_{T,B}}\bigl[c_{F,B} T_r(T_r+1)+
(1-c_{F,B})I(I+1) + (\bar{c}_{F,B}-c_{F,B})T(T+1) \bigr], \eqno(13) $$
where $I$ is the isospin of the $BS$, $T_r$ is the quantity analogous to the
"right" isospin $T_r$ in the collective coordinates approach \cite{3,11,6},
and $\vec{T_r}=\vec{I}^{bf}+\vec{T}$.
$$    c_{F,B}=1-\frac{\Theta_{T,B}}{2\Theta_{F,B}\mu_{F,B}}(\mu_{F,B}-1), 
\qquad
\bar{c}_{F,B}=1-\frac{\Theta_{T,B}}{\Theta_{F,B}(\mu_{F,B})^2}(\mu_{F,B}-1).
 \eqno(14)$$ 
In the rigid oscillator model the states predicted are not identified
with definite $SU(3)$ or $SU(4)$ representations. However, it can be done,
as it was shown in \cite{10}.  The quantization
condition $(p+2q)/3=B$ \cite{3} for arbitrary $N_c$ is changed to 
$(p+2q)=N_cB+3n_{q\bar{q}}$. For example, the state with $c=2$, $I=0$ and
$n_{q\bar{q}}=0$ should belong to the $27-$ plet of $(u,d,c)$ $SU(3)$ group, 
if $N_c=3$, see also \cite{10}. 
For $27$-plet of dibaryons $T_r=1$, for antidecuplet $T_r=0$. 
For $\bar{35}$-plet
of tribaryons $T_r=1/2$, for arbitrary $(p,q)$ irrep which the $BS$ belongs to
$T_r=p/2$ if $n_{q\bar{q}}=0$. $I$ and $T$ take the lowest possible values,
$0$ or $1/2$ in our case.
If $\Theta_F \rightarrow \infty$ Eq. $(13)$ goes
 over into the expression obtained for axially symmetrical $BS$ in 
collective coordinate approach \cite{11}, in realistic case with 
$\Theta_T/\Theta_F \simeq 2.7$ the structure of $(13)$ is more complicated.

The quantum correction due to usual space rotations, also of the order of
$1/N_c$ is exactly of the same form as obtained in \cite{11}, see
\cite{9,10}. 
The binding energies shown in Table 1 are defined relative to the decay
into $B$ baryons, nucleons or flavored hyperons. The binding energy, e.g.
of $B=4$ state relative to $2$ dibaryons will be smaller or negative.
Since we are interested in the lowest energy states we discuss here the
baryonic systems with the lowest allowed angular momentum, $J=0$ for $B=2$, $4$,
and $J=3/2$ for $B=3$. The latter value is due to the constraint because of
symmetry properties of the configuration. The value $J=1/2$ is allowed for
the configuration found in \cite{15}. 
 
For $B=3$ and $4$ toroidal configurations we used here do not correspond
to the minimum of static energy, but only for such configurations the
necessary quantities, $\Theta_{F,B}$, $\Gamma_B$ are known. For $B=3$ the toroidal
configuration does not differ much in energy from the tetrahedral one
which is known to be the configuration of minimal energy \cite{15,16}.
(The masses of stranglets obtained from bound skyrmions with $B$ up to $17$
\cite{16} have been estimated recently in \cite{17} in the bound state
soliton model.)
For $B=4$ the difference is large,  $\sim 300 Mev$ in energy. However,
it would be incorrect to decrease all $B=4$ energies by $300 Mev$ and
increase the binding energies, because other characteristics of solitons and,
therefore, the excitation energies $\omega_c$ and $\omega_b$ also change.
Some reasonable extrapolation for $B=4$ is shown in Table 1.

\vspace{3mm}
\begin{center}
\begin{tabular}{|l|l|l|l|l|l|l|l|l|l|l|l|l|l|}
\hline
 $B$  &$M_{cl,B}$& $\Theta_{F,B}$ & $\Theta_{T,B}$&$\Theta_{J,B}$ &$\Gamma_B$ 
&$\omega_s$&$\omega_c$& $\omega_b$ &$\epsilon_{s=-2}$& $\epsilon_{c=1}$ &
 $\epsilon _{c=2}$ & $\epsilon_{b=-1}$ & $\epsilon_{b=-2}$   \\
\hline
$1$&$0.865$&$1.86$&$5.14$&$5.14$&$3.98$&$0.200$&$1.18$&$3.66$&---&---&---
&---&--- \\
\hline
$2$&$1.656$&$3.79$&$10.55$&$16.45$&$7.80$&$0.196$&$1.15$&$3.62$&$0.096$
&$0.16$&$0.15$&$0.18$&$0.19$\\
\hline
$3$&$2.523$&$6.16$&$16.85$&$37.85$&$12.85$&$0.205$&$1.17$&$3.63$&$0.12$
&$0.22$&$0.23$&$0.27$&$0.27$ \\
\hline
$4$&$3.446$&$8.84$&$23.65$&$72.5$&$18.80$&$0.215$&$1.19$& $3.68$&$0.18$
&$0.23$&$0.21$&$0.25$&$0.25$\\
\hline
$ 4^*$&$3.140$&--- &--- &---&---&$0.196$&$1.15$ &$3.62$&$0.52$
&$0.58$ &$0.61$ &$0.60$&$0.65$ \\
\hline
\end{tabular}
\end{center}
\vspace{2mm}

{\baselineskip=10pt 
\tenrm
{\bf Table 1.} The static characteristics of the $B=1$ hedgehog and toroidal 
solitons with $B=2,3,4$ \cite{14}:
$M_{cl,B}$ in $Gev$, moments of inertia $\Theta_{F,B}=\Theta_c=\Theta_b$, 
$\Theta_T$, $\Theta_J$  and $\Gamma$ in $Gev^{-1}$.
The excitation frequences $\omega_{s,c,b}$ - in $Gev$. The binding energies 
(in $Gev$) of  baryonic systems with $B=2,3,4$, $s=-2$ $(\epsilon_{s=-2})$, 
charm $c=1,2$ ($\epsilon_{c=1,2}$) and bottom $b=-1,-2$ ($\epsilon_{b=-1,-2}$) 
 are shown. The parameters of the 
model $F_{\pi}=108 Mev$, $e=4.84$ \cite{3}. The line $B=4^*$ shows the binding
energies for $B=4$ configuration found in \cite{15,16} with extrapolation 
$\omega_{B=4}=\omega_{B=2}$. The uncertainty of these estimates within our
choice of the model and configurations is $\sim 0.02 Gev$.}

\vspace{3mm}
\elevenrm
The binding energy of the deuteron-like state within this approach is $0.16$
$Gev$, the binding energy of the $NN$ scattering atate with isospin $T=1$
is $0.127$ $ Gev$. If we assume that the nonzero modes quantum corrections -
due to vibration and breathing modes, as well as 1-loop corrections - are
approximately the same for all states with the same $B$ then we should
renormalize the energy of each state obtained in this way, i.e. we should 
subtract just $0.127 Gev$ from the binding energies shown in Table 1 (the
energy of the virtual level, $0.067$ $Mev$, is negligible, of course).
Renormalization of this type was done also previously in \cite{11}.
It is clear that after this renormalization the state with strangeness $S=-2$
becomes unbound, $\sim 0.03$ $Gev$ above the threshold, but the states with
charm or bottom remain bound. These estimates are crude, of course, because the
binding energy of the deuteron is about $\sim 0.03$ $Gev$ within the same 
approach. However, they show clearly that $BS$ with charm and bottom have
more chances to be bound relative to strong interactions than strange $BS$.\\
\section{Symmetry breaking in flavor decay constants}
 The estimates of binding energy were made in \cite{9,10} with the constant
$F_K=F_\pi$, and similar assumption was made above to get the results shown
in Table 1. 
However, symmetry breaking takes place not only in the masses of baryons, but 
also in decay constants, since generally $F_F$ is different from $F_\pi$.
As it was shown by Riska and Scoccola \cite{18}, the large enough values 
of $F_F$ allow to remove the apparent overbinding of heavy mesons by
$SU(2)$ solitons which is characteristic for the bound state models
with $F_F=F_\pi$.

For kaons, $F_K=1.22 F_\pi$, and this is important for the 
description of mass
differences inside the octet and decuplet of baryons.
Therefore, it seems necessary to investigate the binding energies of $BS$
also for the case of this kind of flavor symmetry breaking ($FSB$).

The mass term in the Lagrangian $(3)$ should be changed to
$$ L_M = -(m_F^2 F_F^2/F_\pi^2  - m_\pi^2) \Gamma_B. \eqno (15) $$
It is assumed that the pion mass term is included into the classical mass
of the soliton $M_{cl}$.
Corresponding change should be made in the Hamiltonian $(7)$ also.
The second order term in the Lagrangian should be modified also, but
the corresponding contribution to the mass or energy is much smaller
than that given by $(15)$ because it is proportional to $cos f$ in the
integrand and cancellations take place in the integration over $d^3r$,
see \cite{11,20}.

It is not difficult to calculate the "flavor" moment of inertia in the case
of $FSB$:
$$ \Theta_{F,B} \rightarrow \Theta_{F,B}+(F_F^2/F_\pi^2-1) \Gamma_B/4 \eqno (16) $$
The modified inertia $\Theta_F$ can be calculated easily according to $(16)$
for any $F_F$ since $\Theta_F$ and $\Gamma_B$ are known. 

The following expression can be obtained for the excitation energy:
$$\omega_{F,B} \simeq \frac{m_F}{[1 + 4 \Theta_{F,B}^{Sk}/(r_F^2 \Gamma_B)]
^{1/2}} - \frac{N_cB}{8 \Theta_{F,B}}. \eqno (17) $$
Here $r_F=F_F/F_\pi$, $\Theta_{F,B}^{Sk}$ is the Skyrme term 
contribution to the moment of inertia, so we can write:
$$ \Theta_{F,B} = \Theta_{F,B}^{Sk} + r_F^2 \Gamma_B/4. \eqno (18) $$
For our choice of the model parameters $\Theta_{F,B}^{Sk} $ is
a bit smaller than  $\Gamma_B/4$.
\vspace{3mm}
\begin{center}
\begin{tabular}{|l|l|l|l|l|l|l|l|l|l|l|l|l|}
\hline
 $B$  & $\Theta_{s,B}$ & $\Theta_{c,B}$&$\Theta_{b,B}$ 
&$\omega_s$&$\omega_c$& $\omega_b$&$\bar{c}_{s,B}$&$\epsilon_{s=-2}$& 
$\bar{c}_{c,B}$ & $\epsilon _{c=2}$ & $\bar{c}_{b,B}$ & $\epsilon_{b=-2}$  \\
\hline
$1$ &$2.346$&$3.104$&$4.845$&$0.255$&$1.466$&$4.708$&$0.480$&---&$0.883$&---
&$0.980$&--- \\
\hline
$2$ &$4.742$&$6.228$&$9.640$&$0.250$&$1.447$&$4.671$&$0.472$&$0.11$
&$0.880$&$0.16$&$0.982$&$0.20$\\
\hline
$3$ &$7.729$&$10.18$&$15.80$&$0.260$&$1.464$&$4.684$&$0.498$&$0.15$
&$0.892$&$0.22$&$0.984$&$0.28$ \\
\hline
$4$ &$11.13$&$14.71$&$22.94$&$0.270$&$1.481$&$4.713$& $0.528$&$0.20$
&$0.903$&$0.26$&$0.986$&$0.30$\\
\hline
$ 4^*$ &--- &--- &--- &$0.250$&$1.447$ &$4.671$&$-$
&$0.52$ &---  &$0.60$&---&$0.66$  \\
\hline
\end{tabular}
\end{center}
\vspace{1mm}

{\baselineskip=10pt 
\tenrm
{\bf Table 2.} The flavor inertia $\Theta_{F,B}$ (in $Gev$) for $F_K/F_\pi=1.22$, $F_c/F_\pi=1.5$,
$F_b/F_\pi=2$. 
Excitation frequences and some binding energies for $|F|=2$ baryonic 
          systems with flavor symmetry breaking in the constants $F_F$
           - in $Gev$.
          The quantities $\bar{c}_{F,B}$ defining the suppression
of zero-modes quantum corrections are shown also.}
\vspace{3mm}

We took here $r_c=F_D/F_\pi=1.5$ and $r_b=F_b/F_\pi=2$ because these values
of decay constants ratios allow to get the masses of lowest baryons 
$\Lambda_c$ and $\Lambda_b$ not very far from the observed values. 
$r_c$ does not contradict to the experimental restriction.
As it can be seen from Table 2 the flavor symmetry violation in meson
decay constants leads to moderate increase of the binding energies of $BS$
with different flavors, and the tendency of increase of $\epsilon_F$ with
increasing mass of flavor becomes more striking. 

It is of interest to look at the case of large $FSB$, when the ratio
$r_F=F_F/F_\pi \gg 1$. In this case it is easy to find out what is 
the effective potential
for the heavy meson in the $SU(2)$ soliton background.
The following expression can be obtained:
$$\omega_{F,B} \simeq m_F [1-2\Theta_{F,B}^{Sk}/(\Gamma_B r_F^2)] -
m_\pi^2/(2m_Fr_F^2) -N_cB/(2\Gamma_B r_F^2) \eqno(19) $$
This defines the average potential of the heavy meson bound by 
$SU(2)$ soliton. Obviously, $\omega_{F,B} \to m_F$ when $r_F \to \infty$.\\
\section{Conclusions}
 To conclude, we estimated the binding energies of dibaryons,
tribaryons and tetrabaryons with nonzero charm and bottom. For the top 
quantum number similar results can be obtained, but
the spectroscopy of mesons, baryons and baryonic systems with t-number 
will not be available, probably, 
because of the large width of the top quark. 

For $r_t=F_t/F_\pi=2$ the $\omega_t -s$
are about $157-158$ $Gev$ and the binding energies $\sim 1 Gev$.
To obtain the excitation energies $\omega \simeq m_t=175$ $Gev$ and the binding
energies of the same order as we have, e.g. for strange $BS$, we should take
$r_t \sim \sqrt{m_t}$. For example, for $r_t=17$ $\omega \simeq 174.7$ $Gev$
and $\epsilon_{t=2} \simeq 0.2 Gev$. However, such big value of $r_t$ seems
to be unrealistic. Anyway, it may be of interest that for very large mass 
of the "flavored" meson the scale of the binding energies of $BS$ is 
connected with the scale of the ratio $F_F/F_\pi$.

Since the binding energies for strangeness, charm or bottom increase with 
increasing mass of the flavor for realistic values of constant $F$,
the charmed and bottomed baryonic systems have more chances to be bound 
than strange $BS$. This is in agreement with the experimental fact 
that the difference of masses of $\Lambda_F$ baryons and corresponding
pseudoscalar mesons, $P_F$ ($K$, $D$ or $B$) decreases with increasing
mass of flavor, just indicating that the binding energy of flavored
quark in the baryon increases in comparison with that in meson with
increasing $m_F$. 
 Nonzero quantum corrections to the energy of charmed (bottomed) 
baryonic systems are expected to be smaller in comparison with strange 
baryonic systems, because of the greater mass of charmed (bottomed) 
quarks or mesons.

The apparent drawback of the approach exploited in the present paper is
that the motion of the system into the "charm" or "bottom" direction is
considered independently from other motions. Consideration of
the $BS$ with "mixed" flavors is  possible, in principle, but it demands more
complicated treatment.

The collective coordinates approach with 
the rigid or soft rotator variant of the model usually gives the 
masses of baryons considerably greater than the Kaplan-Klebanov-Westerberg
model we used here, if the Casimir energies are not taken into account
 \cite{3,11,18}. One of the sources of this difference is the presence
of the zero-modes contribution in the rotation energy of the order of
$N_c/ \Theta_F$, see $(11)$ \cite{13,11,18}, which is absent in the oscillator 
model. As it was shown recently by Walliser for the $B=1$ sector \cite{13} this 
large contribution is cancelled almost completely by the 1-loop correction --- 
zero-point Casimir energy which is of the same order of magnitude, $N_c^0$ 
\cite{19}. Anyway, since both approaches have led to similar results in
the case of strange baryonic systems, we may expect the same 
for the case of charmlets and bottomlets, so, our results should be valid 
qualitatively, at least.

The threshold for the charm production on a free nucleon is about $12 Gev$,
for double charm - about $25.2 Gev$. However, for nuclei as a target the
thresholds are much lower due to two-step processes with mesons in
intermediate states and due to normal Fermi-motion of nucleons inside the 
target nucleus (see, e.g. \cite{22}).
Therefore, production of states with $c=1$ and even $c=2$ will be available
on accelerators like future Japan Hadron Facility (energy $\sim 50 Gev$),
the subthreshold production of $b=-1$ systems on nuclei with small probability 
also will be possible.

When the present paper was almost completed I have found a paper \cite{23}
where the charmed few baryon systems have been considered within more
conventional potential approach. The $c=1$ system with $B=3$ was found to be
not bound, probably (the binding energy is of the order of $1 Mev$),
The $B=4$ system was found to be bound with the binding energy not 
exceeding $\sim 10 Mev$.

I am indebted to H.Walliser for useful discussions of the skyrmions
quantization at arbitrary $N_c$ and $N_F$, and for valuable remarks.
I appreciate also critical remarks by A.Penin.
\vglue 0.4cm
{\elevenbf\noindent References}
\vglue 0.2cm

\end{document}